\documentclass[12pt,american,pra,reprint,onecolumn]{revtex4-1}
\usepackage[T1]{fontenc}
\usepackage[latin9]{inputenc}
\usepackage{babel}
\usepackage{units}
\usepackage{amsthm}
\usepackage{amsmath}
\usepackage{amssymb}
\usepackage[unicode=true,pdfusetitle,
 bookmarks=true,bookmarksnumbered=false,bookmarksopen=false,
 breaklinks=false,pdfborder={0 0 1},backref=false,colorlinks=false]
 {hyperref}
\usepackage{breakurl}

\makeatletter
\@ifundefined{textcolor}{}
{%
 \definecolor{BLACK}{gray}{0}
 \definecolor{WHITE}{gray}{1}
 \definecolor{RED}{rgb}{1,0,0}
 \definecolor{GREEN}{rgb}{0,1,0}
 \definecolor{BLUE}{rgb}{0,0,1}
 \definecolor{CYAN}{cmyk}{1,0,0,0}
 \definecolor{MAGENTA}{cmyk}{0,1,0,0}
 \definecolor{YELLOW}{cmyk}{0,0,1,0}
}
\theoremstyle{plain}
\newtheorem{thm}{\protect\theoremname}
  \theoremstyle{remark}
  \newtheorem{rem}[thm]{\protect\remarkname}

\makeatother

  \providecommand{\remarkname}{Remark}
\providecommand{\theoremname}{Theorem}

\begin{document}

\title{A paradox about an atom and a photon}

\author{Carlo Maria Scandolo}

\affiliation{Università degli Studi di Padova, Scuola Galileiana di Studi Superiori}

\email{carlomaria.scandolo@studenti.unipd.it}

\begin{abstract}
In this article we propose a new relativistic paradox concerning the
absorption of a photon by a hydrogen atom. We show that the actual
cause of the paradox is one of the hypotheses of Bohr model; therefore,
in order to solve the paradox, we have to move away from Bohr model.
Our analysis is carried out only in the special relativistic framework,
so we are not interested in giving a full quantum mechanical treatment
of the problem. We derive some expressions for emission and absorption
of photons by atoms, which are in perfect agreement with special relativity,
although comparable to the classical Bohr formula with an excellent
degree of approximation. Quite interestingly, these expressions are
no more invariant under a global shift of energy levels, showing a
breaking of classical ``gauge invariance'' of energy. We stress
that, to the best of our knowledge, the present approach has never
been considered in literature. At the end we will be able to solve
the proposed paradox.
\end{abstract}



\maketitle

\section{Introduction}

So far the special theory of relativity has proven to be one of the
most fruitful theories both for the physical consequences and the
great conceptual depth, which has allowed us to get a better insight
about Nature. Moreover, special relativity is often counter-intuitive
to common sense, given our experience in a world in which Galilean
relativity is in force up to an excellent degree of approximation.

Just because of the conceptual relevance of special relativity and
of its counter-intuitiveness, the use of paradoxes is one of the essential
tools to gain a deeper comprehension of the theory itself and of its
applications. It is enough to recall the famous twin paradox and ladder
paradox, see \cite{Rindler}.

In every special relativistic paradox there is a physical phenomenon
described by (at least) two different observers, linked to as many
inertial reference frames. The paradox consists in the fact that these
different observers give contradictory descriptions of the phenomenon,
and this fact is in contrast with relativity principle. Actually,
at least one of the observers makes a mistake in his assumptions or
when trying to apply physical laws to the phenomenon in his reference
frame, so the violation of relativity principle is only seeming.

Therefore, when solving a paradox, there are some goals to achieve.
In particular we must understand 
\begin{enumerate}
\item which of the observers is (are) wrong 
\item what actually happens 
\item \label{enu:why is wrong}why that (those) observer(s) is (are) wrong 
\end{enumerate}
According to this working scheme, the great conceptual importance
of solving a paradox is apparent. In particular, through the treatment
of point~\ref{enu:why is wrong}, one has the opportunity to detect
some insidious errors in the application of the theory, thus allowing
a deeper and deeper understanding of the theory. Wheeler \cite{Wheeler}
himself stated the importance of paradoxes in physical progress and
understanding.

In view of the importance of the paradox tool, in this article we
will start from a new relativistic paradox, which, from what we know,
has never been presented before in literature. In this paradox, a
hydrogen atom interacts with a photon, and different observers come
to different conclusions about the absorption of the photon.

Emission and absorption of photons by atoms are well known phenomena
and they may look rather simple. However, in this article we will
show that they are not as trivial as they might appear at first glance.

In particular, Bohr model of hydrogen atom \cite{Bohr,Bohm,Beiser}
establishes that an atom can absorb or emit photons with energies
equal to energy gaps between atomic levels. We will show that this
assumption, which is the core of this paradox, leads to inconsistencies
from a relativistic perspective.

In this way, we will be able to show once more that Bohr model of
hydrogen atom is definitely ruled out because of its incompatibility
with special relativity. This conclusion should not surprise, because
it is a well-known and established fact, but it is interesting to
explore the origin of this contradiction, and we will do that by means
of a paradox. We would like to stress that such an approach to the
incompatibility of Bohr model with special relativity has never been
considered in literature for all we know.

Therefore this paradox will be an opportunity to discuss the role
of photons in atomic excitation and disexcitation from a purely relativistic
perspective.

Since we are mainly interested in special relativity, we will not
deal with the problem from a quantum mechanical point of view.

\section{\label{sec:Statement-of-the}Statement of the paradox}

Now we are ready to state the paradox, which is the core of this article.

Let $K$ be an inertial frame (the laboratory frame). A hydrogen atom
in its ground state is moving in $K$ along the positive direction
of the $x$ axis with speed $v$. A photon with energy $E_{0}$ is
moving in the opposite direction. Let $E_{0}$ be less than the energy
necessary in order to get the atom to its first excited state. An
observer in $K$ states that the photon will not be absorbed by the
atom.

Let $K'$ be a reference frame which is stationary relative to the
hydrogen atom, viz. it is moving with speed $v$ relative to $K$.
Due to Doppler effect, an observer in $K'$ sees the photon with a
higher frequency, and therefore with a higher energy. For a suitable
speed $v$, it is possible in $K'$ to see the photon just with the
energy of the first quantum jump. In this case, an observer in $K'$
will state that the photon will be absorbed.

Let us assume, for the sake of simplicity, that the spatial axes of
$K'$ are parallel to those of $K$ and that time flows in the same
direction both in $K$ and $K'$.

With a careful analysis we can isolate two tacit assumptions in this
paradox: 
\begin{enumerate}
\item the energies of the various quantum jumps are the same in every inertial
frame
\item a photon can be absorbed only if it has the energy of a quantum jump
(Bohr hypothesis) 
\end{enumerate}
Now we will examine these two assumptions critically. Throughout this
article we will adopt the metric signature $+---$ for Minkowski metric
tensor $\eta_{\mu\nu}$.

\section{Are energy levels invariant?}

For the purposes of this article, we will not need to deal with the
relativistic quantum mechanical treatment of the hydrogen atom, even
though we will examine the paradox in the special relativistic framework.
For the sake of simplicity, we will follow the ``classical'' quantum
mechanical paradigm.

From this treatment \cite{Cohen} of the hydrogen atom, it is well
known that the Hamiltonian eigenvalues (from now on called simply
``energy levels'') are 
\begin{equation}
E_{n}=-\frac{1}{2}\mu c^{2}\alpha^{2}\frac{1}{n^{2}},\label{eq:energy levels}
\end{equation}
where $\mu$ is the reduced mass of the system, $\alpha$ is the fine-structure
constant, and $n$ is a strictly positive integer. A so-called ``quantum
jump'' is nothing but the energy gap between two energy levels of
the atom.

Now we have to deal with the energy levels of eq.~\eqref{eq:energy levels}
in the relativistic framework. It is then convenient to introduce
the effective mass $m_{n}$ of the hydrogen atom in its $n$-th energy
level. The effective mass is defined as 
\begin{equation}
m_{n}:=M+\frac{E_{n}}{c^{2}}=M-\frac{1}{2}\mu\alpha^{2}\frac{1}{n^{2}},\label{eq:effective mass}
\end{equation}
where $M$ is the total mass of the electron-proton system. A quick
check assures us that $m_{n}>0$ for every positive integer $n$;
so this quantity is always well-defined.

A question now arises: are the energy levels of eq.~\eqref{eq:energy levels}
invariant? Or are they related to a particular reference frame?

From the quantum mechanical treatment of the hydrogen atom, energy
levels are naturally related to the inertial frame which is stationary
with respect to the center of mass of the atom. That frame is $K'$.

In order to find the energy levels of the hydrogen atom in another
frame $K$, moving with constant speed $v$ with respect to $K'$,
we have to transform the 4-momentum of the atom, where the effective
mass behaves for all practical purposes exactly as a rest mass.

Thus, we have that the total energy of the hydrogen atom in $K$ is
\[
E=\gamma Mc^{2}-\frac{1}{2}\gamma\mu c^{2}\alpha^{2}\frac{1}{n^{2}},
\]
where, as usual, $\gamma=\left(1-\beta^{2}\right)^{-1/2}$ and $\beta=v/c$.

There is a correction term $-\gamma\mu c^{2}\alpha^{2}/\left(2n^{2}\right)$
to the energy $\gamma Mc^{2}$ of the compound system given by proton
and electron, because of the binding energy of the electron. Therefore,
the energy levels of the moving atom are \emph{not} the same as those
of the stationary atom; more precisely they are 
\begin{equation}
E_{n}\left(v\right)=-\frac{1}{2}\gamma\mu c^{2}\alpha^{2}\frac{1}{n^{2}}.\label{eq:energy levels in motion}
\end{equation}
We notice the presence of the multiplicative corrective term $\gamma$
to the energy levels of eq.~\eqref{eq:energy levels}, which are
seen in the center of mass reference frame ($K'$). Since the correction
is multiplicative, and not additive, quantum jumps in $K$ are \emph{not}
the same as in $K'$. 

We have just got the main result of this section: the energy levels
of a hydrogen atom observed in a given inertial frame depend on the
motion state of the atom with respect to the given inertial frame.
Therefore, the assumption of invariance of energy levels and quantum
jumps is \emph{wrong}.

\section{Are photon energies the same as those of quantum jumps?}

Now let us analyze the issue of photon absorption and emission by
atoms. From Bohr model Ritz-Rydberg formula follows easily, see \cite{Bohm,Beiser}.
Ritz-Rydberg formula relates the emitted wavelengths $\lambda$ to
the quantum transition between two different energy levels. 
\[
\frac{1}{\lambda}=\frac{\mu e^{4}}{8\varepsilon_{0}^{2}h^{3}c}\left(\frac{1}{n^{2}}-\frac{1}{m^{2}}\right),
\]
where $m>n$. Here $m$ and $n$ are positive integers respectively
labeling the energy levels $E_{n}$ and $E_{m}$ of eq.~\eqref{eq:energy levels}.
The energy $E_{m,n}$ of the emitted photons in the transition from
the $m$-th to the $n$-th energy level is given by $E_{m,n}=hc/\lambda$,
whence 
\begin{equation}
E_{m,n}=E_{m}-E_{n}.\label{eq:emitted photon Rydberg}
\end{equation}

In this way we are naturally led to assume that emitted (and likewise
absorbed) photons have energies corresponding to those of quantum
jumps.

Nevertheless, in this treatment we have completely neglected linear
momentum and energy conservation laws. Actually, we have to take into
account 4-momentum conservation for the system made up of the atom
and the photon, since the interaction of the hydrogen atom with the
photon does not involve external 4-forces. Total 4-momentum conservation
law thus holds both for emission and absorption of photons.

However, the analysis of photon emission is perhaps conceptually simpler
than the absorption case, so we will begin analyzing the emission
case.

\subsection{The emission case}

\subsubsection{\label{sub:Emission-by-an}Emission by an atom at rest}

Let us consider a hydrogen atom in its $m$-th energy level, and let
us suppose that the atom is stationary relative to the laboratory
reference frame $K$. From now on, we will analyze the physical situation
in this frame. The atom returns to a lower energy level $E_{n}$ ($n<m$)
through the emission of a photon.

Now we will apply 4-momentum conservation law to the emission process,
using the effective mass, defined in eq.~\eqref{eq:effective mass}.
Let us suppose that the photon is emitted in the laboratory frame
along the direction defined by the unit vector $\vec{n}$.

From 4-momentum conservation, and from the mass-shell relation with
the effective mass, we get an equation for $E_{m,n}$, which yields
(cf.~\cite{Barone}) 
\begin{equation}
E_{m,n}=\left[1-\frac{E_{m}-E_{n}}{2\left(Mc^{2}+E_{m}\right)}\right]\left(E_{m}-E_{n}\right).\label{eq:emitted photon energy levels}
\end{equation}
Notice that the result of eq.~\eqref{eq:emitted photon energy levels}
does not depend on the direction of emission $\vec{n}$ in the laboratory
frame, according to space isotropy.

However, the most important fact is that there is a corrective term,
so the square bracket is always strictly less than 1, provided $m\neq n$,
as it is the case. We conclude that $E_{m,n}<E_{m}-E_{n}$.
\begin{rem}
From 4-momentum conservation it easy to see that a hydrogen atom cannot
jump to a lower energy level without the emission of a photon.

In fact, from 3-momentum conservation, if a disexcitation without
a photon emission were possible, the atom would be at rest both in
the initial and in the final state. Therefore, from energy conservation,
we would have $m_{n}=m_{m}$, which is impossible because $m\neq n$.

Since 4-momentum conservation is a covariant law, this result, derived
for an atom at rest, holds also for an atom moving at constant velocity. 
\end{rem}
Now we want to investigate if 4-momentum conservation provides us
with some selection rule on $m$ and $n$. Therefore we ask ourselves
if the square bracket of eq.~\eqref{eq:emitted photon energy levels}
may vanish. In that case there would be no emission at all. However,
it is not difficult to show that this fact can never happen, so 4-momentum
conservation imposes no selection rule.

Let us evaluate the result of eq.~\eqref{eq:emitted photon energy levels}
also from a numerical point of view. In a hydrogen atom, $Mc^{2}\sim\unit[1]{GeV}$,
whereas $\left|E_{n}\right|<\unit[13.6]{eV}$, so the second term
in the square bracket is negligible. Therefore, up to an excellent
degree of approximation, eq.~\eqref{eq:emitted photon energy levels}
drops to eq.~\eqref{eq:emitted photon Rydberg}. Nevertheless, from
a rigorous point of view, the emitted photon does \emph{not} have
the energy of the quantum jump, but a lower energy, due to the fact
that the atom recoils. In the continuation of this article, we will
always proceed in a rigorous manner, so we will not neglect the atom
recoil, and we will always use eq.~\eqref{eq:emitted photon energy levels},
which is in agreement with 4-momentum conservation, unlike eq.~\eqref{eq:emitted photon Rydberg}.

So far we have analyzed the emission of a photon in the laboratory
reference frame. Now let us move to consider the frame which is stationary
relative to the hydrogen atom after the emission of the photon. What
energy of the photon will an observer in this frame see?

Let $u^{\mu}$ be the recoil 4-velocity of the atom; and here we take
$u^{\mu}$ to be dimensionless. Using tetrad formalism, it is now
easy to calculate the energy $E'_{m,n}$ of the emitted photon according
to an observer co-moving with the recoiling atom. $E'_{m,n}$ is given
by 
\[
E'_{m,n}=cp_{0}^{\mu}u_{\mu},
\]
where $p_{0}^{\mu}$ is the emitted photon 4-momentum. After some
passages, we come to an equation very close to eq.~\eqref{eq:emitted photon energy levels},
which is 
\[
E'_{m,n}=\left[1+\frac{E_{m}-E_{n}}{2\left(Mc^{2}+E_{n}\right)}\right]\left(E_{m}-E_{n}\right).
\]
In this equation there is a plus sign, so, according to the co-moving
observer, the energy of the emitted photon is greater than the energy
gap between $E_{m}$ and $E_{n}$.

We conclude that neither in the recoiling atom co-moving frame the
photon is emitted with the same energy as the energy gap between the
two energy levels.

Therefore, at least in the emission case from an atom at rest, the
assumption that an emitted photon has an energy equal to that of the
quantum jump proves to be \emph{wrong}.

\subsubsection{Emission by a moving atom}

Let us examine the case of photon emission by a hydrogen atom moving
at constant speed $v$ along the $x$ axis of the laboratory reference
frame $K$.

We might think that the relation given by eq.~\eqref{eq:emitted photon energy levels}
holds in this case too, provided we replace $E_{n}$ and $E_{m}$
with the energy levels $E_{n}\left(v\right)$ and $E_{m}\left(v\right)$
of a moving atom, given by eq.~\eqref{eq:energy levels in motion},
and we put $\gamma M$ in place of $M$. In this way we would get
\begin{equation}
E_{m,n}=\gamma E_{m,n}^{*},\label{eq:false energy in-flight emission}
\end{equation}
where $E_{m,n}^{*}$ is the energy of the emitted photon when the
atom is at rest, see eq.~\eqref{eq:emitted photon energy levels}.
Nevertheless, we can immediately rule out this result by a general
argument. The fact that the atom is moving along the $x$ axis breaks
space isotropy, so an equation describing photon emission has to depend
on the direction of emission of the photon. The result in eq.~\eqref{eq:false energy in-flight emission}
is independent of the direction of emission, therefore it has to be
\emph{ruled out}.

Let us find the correct expression. Let $\vartheta$ be the angle
between the direction of emission, given by the unit vector $\vec{n}$,
and the $x$ axis of $K$. We can transform the photon 4-momentum
from the the laboratory frame to the co-moving frame. We obtain 
\begin{equation}
E_{m,n}=\frac{E_{m,n}^{*}}{\gamma\left(1-\beta\cos\vartheta\right)},\label{eq:true in-flight energy emission}
\end{equation}
where $E_{m,n}^{*}$ is again given by eq.~\eqref{eq:emitted photon energy levels}.
Notice that eq.~\eqref{eq:true in-flight energy emission} correctly
depends on the direction of the emission of the photon through the
angle $\vartheta$.

In particular, one has the minimum energy for the emitted photon when
$\vartheta=\pi$, that is when the photon is emitted in the opposite
direction to the direction of motion; the maximum energy $E_{\mathrm{max}}$
of the emitted photon is instead attained when $\vartheta=0$, that
is when the photon is emitted in a parallel direction to the direction
of motion.

It is not difficult to prove that for a speed $v$ such that 
\[
\frac{1-A^{2}}{1+A^{2}}<\beta<1,
\]
where $A$ is the square bracket of eq.~\eqref{eq:emitted photon energy levels},
we have $E_{\mathrm{max}}>E_{m}-E_{n}$. In this way the energy of
the emitted photon is greater than the energy gap between $E_{m}$
and $E_{n}$. In particular, for $\beta=\left(1-A^{2}\right)/\left(1+A^{2}\right)$,
we drop again to eq.~\eqref{eq:emitted photon Rydberg} and Ritz-Rydberg
formula: in this particular case the photon has the same energy as
the gap $E_{m}-E_{n}$.

\subsection{The absorption case}

\subsubsection{Absorption by an atom at rest}

Now, let us move to examine a situation which is closer to our paradox.
Let us consider a hydrogen atom in its $n$-th energy level at rest
in the laboratory frame $K$. A photon with energy $E_{0}$ in $K$
is moving towards the atom along the direction given by the unit vector
$\vec{n}$ in the laboratory frame. What energy does the photon need
to have in order to get the atom to its $m$-th energy level ($m>n$)?

From the treatment of the emission case, we may figure that this energy
is not equal to the energy gap between the $m$-th and the $n$-th
energy level. Now we will analyze the absorption case in a similar
fashion to what we have done for the emission case. 

By means of a perfectly analogous procedure, we finally get (cf.~\cite{Barone})
\begin{equation}
E_{0}=\left[1+\frac{E_{m}-E_{n}}{2\left(Mc^{2}+E_{n}\right)}\right]\left(E_{m}-E_{n}\right).\label{eq:absorbed photon energy levels}
\end{equation}
Again the result is independent of the direction of the incident photon,
due to space isotropy.

In the absorption case, however, the energy of the absorbed photon
is higher than the energy gap of the corresponding quantum transition,
due to the plus sign in the square bracket.

In this case, an analysis of atomic excitation without photon absorption
is not as trivial as atomic disexcitation without photons, thus we
will defer it until section~\ref{sec:When-the-photon}.

It is, instead, obvious that 4-momentum conservation does not impose
any selection rule also in the absorption process.

Finally, since the estimates in section~\ref{sub:Emission-by-an}
still hold, eq.~\eqref{eq:absorbed photon energy levels} drops to
eq.~\eqref{eq:emitted photon Rydberg} up to an excellent degree
of approximation.

\subsubsection{Absorption by a moving atom}

Now, let us analyze the case of photon absorption by a hydrogen atom
moving with speed $v$ along the $x$ axis of the laboratory frame
$K$. As in the preceding section, a photon is moving along the direction
given by the unit vector $\vec{n}$.

The motion of the atom breaks space isotropy, so, as in the case of
photon emission by a moving atom, an equation like eq.~\eqref{eq:false energy in-flight emission}
is not allowed.

Transforming the photon 4-momentum, we get exactly eq.~\eqref{eq:true in-flight energy emission},
but now $E_{m,n}^{*}$ is given by eq.~\eqref{eq:absorbed photon energy levels}.
\begin{equation}
E_{m,n}=\frac{E_{m,n}^{*}}{\gamma\left(1-\beta\cos\vartheta\right)}\label{eq:in-flight energy absorption}
\end{equation}

It is possible to show that the minimum energy $E_{\mathrm{min}}$
(when $\vartheta=\pi$), for a speed $v$ such that 
\[
\frac{B^{2}-1}{B^{2}+1}<\beta<1,
\]
where $B$ is the square bracket of eq.~\eqref{eq:absorbed photon energy levels},
is lower than the energy gap of the quantum transition, $E_{m}-E_{n}$.
In particular, for $\beta=\left(B^{2}-1\right)/\left(B^{2}+1\right)$,
we drop again to eq.~\eqref{eq:emitted photon Rydberg} and to Ritz-Rydberg
formula: in this particular case the photon energy is the same as
$E_{m}-E_{n}$.

\section{\label{sec:When-the-photon}When the photon is not absorbed\ldots{}}

In the previous section we analyzed the case when the photon has the
right energy to be absorbed. What happens if the photon is not absorbed?

We wonder whether a hydrogen atom in its energy level $E_{n}$ can
jump to a higher level $E_{m}$ without a photon absorption. We will
carry out this analysis in the atom rest frame $K$, for the sake
of simplicity, then the result will extend to every inertial frame,
due to covariance.

Let us, then, consider a photon with energy $E_{0}$ moving along
the $x$ axis towards the atom at rest. Let $\varphi$ be the angle
between the scattered photon direction (after interaction) and the
$x$ axis.

In this case, the calculations are not so straightforward, thus we
will give more details in Appendix~\ref{sec:Some-calculations}.
Instead, the treatment in this section will present only the results.

Working out Mandelstam $t$ invariant and imposing energy conservation,
we have that, if excitation takes place, the scattered photon must
have an energy $E_{0,\mathrm{fin}}$, given by 
\begin{equation}
E_{0,\mathrm{fin}}=\frac{\left(m_{n}^{2}-m_{m}^{2}\right)c^{2}+2m_{n}E_{0}}{2\left[E_{0}\left(1-\cos\varphi\right)+m_{n}c^{2}\right]}c^{2},\label{eq:scattered photon energy}
\end{equation}
where $m_{n}$ and $m_{m}$ are the effective masses related to the
energy levels $E_{n}$ and $E_{m}$ respectively.

Clearly it must be $E_{0,\mathrm{fin}}>0$. Imposing this inequality,
we have that the case $E_{0}<E_{\mathrm{abs}}$, where $E_{\mathrm{abs}}$
is the energy a photon must have to be absorbed (see eq.~\eqref{eq:absorbed photon energy levels}),
is ruled out. We conclude that when the photon has a lower energy
than the energy necessary to be absorbed, it cannot excite the atom.

What happens if $E_{0}>E_{\mathrm{abs}}$?

This case seems to be not ruled out, because it implies $E_{0,\mathrm{fin}}>0$
and there is no apparent contradiction. However, we must require that
the recoiling atom has $\gamma\geq1$. Anyway, it is not hard to prove
that this inequality is always fulfilled.

It seems, then, that if $E_{0}>E_{\mathrm{abs}}$, there is no contradiction
in assuming that the atom can jump to an excited state without the
absorption of the photon.

Actually, our treatment cannot exclude absorption completely. In fact,
using 4-momentum conservation, we abandon the attempt of an accurate
description of the interaction between the atom and the photon, treating
the actual interaction as if it were in a sort of ``black box''.
The 4-momentum conservation method is well suited only to describe
the initial state of the system before going into the ``black box''
and the final state of the system after it has come out from the ``black
box''.

So, if we know that in the final state there is an excited atom and
a photon, we cannot conclude that the excitation has occurred without
the intervention of the photon. In fact, in the ``black box'' the
photon can be absorbed, getting the atom from the energy level $E_{n}$
to a higher energy level $E_{m'}$ than $E_{m}$ ($m'>m$), and then
the photon is emitted, getting the atom to the energy level $E_{m}$.
Perhaps several of these absorption-emission processes may take place
in the ``black box''.

To sum up what we have obtained in this section, we can distinguish
two cases when the photon does not have the right energy $E_{\mathrm{abs}}$
to be absorbed. 
\begin{enumerate}
\item $E_{0}<E_{\mathrm{abs}}$: it is impossible to obtain an atomic excitation
to the $m$-th energy level; 
\item $E_{0}>E_{\mathrm{abs}}$: an atomic excitation to the $m$-th energy
level is possible. 
\end{enumerate}

\section{Solution of the paradox}

Now it is time to give the solution to the paradox stated in section~\ref{sec:Statement-of-the}.

In the paradox we are in the absorption case. In the inertial frame
$K$, the hydrogen atom in its ground state is moving with speed $v$,
and the photon is moving in the opposite direction towards the atom.
Therefore, here $\vartheta=\pi$.

The core of this paradox is the fact that it is stated in a vague
way. In fact, the two observers tacitly assume that photons are absorbed
only if they have an energy equal to the energy gap between two energy
levels, independently of the motion state of the atom. Therefore,
in this solution we will interpret the statement ``in $K$ the photon
energy is less than the energy necessary to get the atom to its first
excited state'' as ``in $K$ the photon has an energy which is less
than the energy gap of the first quantum jump''. Likewise, we will
interpret ``in $K'$ the photon has the right energy to be absorbed''
as ``in $K'$ the photon has the energy of the first quantum jump''.

According to the working scheme outlined in the introduction, now
we will give a solution for each of the three points. So, first of
all, let us describe what actually happens.

Let $E_{0}$ be the energy of the incoming photon in $K$, such that
$E_{0}<\Delta E$, where $\Delta E$ is the energy gap between the
first excited level and the ground level.

$E_{0}$ is such that $E'_{0}$, the energy in the co-moving frame
$K'$, is just $\Delta E$. Transforming the photon energy, we have
\[
E_{0}=\frac{\Delta E}{\gamma\left(1+\beta\right)},
\]
and this expression is quite different from the minimum energy expression
in eq.~\eqref{eq:in-flight energy absorption}, because $E_{m,n}^{*}\neq\Delta E$.
Therefore, the photon \emph{cannot} be absorbed in $K$.

The photon \emph{cannot} be absorbed also in $K'$, because it should
have a greater energy than $\Delta E$, as prescribed by eq.~\eqref{eq:absorbed photon energy levels}.
Thus, the photon \emph{cannot} be absorbed in both the frames and
a photon detector will detect the photon after the atomic interaction
both in $K$ and $K'$.

In addition, we can say that the hydrogen atom will not get excited
because of the interaction with the photon, since the energy of the
photon in $K'$, where the atom is at rest, is less than the absorption
energy $E_{\mathrm{abs}}$. This fact, according to section~\ref{sec:When-the-photon},
rules out an atomic excitation to the first excited level.

So, we have understood that the photon is not absorbed both in $K$
and $K'$, so $K$ is right, although this is a mere coincidence,
since both their lines of reasoning are incorrect. In fact, they are
based on two wrong assumptions, those outlined at the end of section~\ref{sec:Statement-of-the}.

Finally, we can explain the error made by $K'$ (and also by $K$)
by his faith in Bohr model, and in the two tacit assumptions of section~\ref{sec:Statement-of-the}.

\section{Conclusions}

In their book \cite{Aharonov} Aharonov and Rohrlich try to classify
the different kinds of paradoxes in physics. They identify three different
classes of paradoxes. 
\begin{description}
\item [{Errors}] The paradox arises from an error in logic or in the understanding
of a particular physical theory. 
\item [{Gaps}] The paradox arises by a flaw in the theory, albeit not a
fatal flaw. It is simply due to a ``gap'' in the theory. 
\item [{Contradictions}] The paradox arises from a fatal flaw, it indicates
that the theory is wrong and it has to be changed. 
\end{description}
According to this classification scheme, the paradox presented in
this article places itself between the error and the gap class. In
fact, the paradox arises from an incorrect interpretation of the absorption
process, but yet the error is due to the use of Bohr model.

In this way, Bohr model, with its assumption that emitted and absorbed
photons have the same energies as those of quantum jumps, gives rise
to a contradiction. Therefore, we must rule out this model.

Instead, special relativity yields a formula for emitted and absorbed
photons between the energy levels $E_{\mathrm{in}}$ and $E_{\mathrm{fin}}$,
which is, for an atom at rest, 
\begin{equation}
E_{\mathrm{ph}}=\left[1+\frac{E_{\mathrm{fin}}-E_{\mathrm{in}}}{2\left(Mc^{2}+E_{\mathrm{in}}\right)}\right]\Delta E,\label{eq:summing up}
\end{equation}
where $E_{\mathrm{ph}}$ is the photon energy and $\Delta E=\left|E_{\mathrm{fin}}-E_{\mathrm{in}}\right|$.
This formula correctly predicts no emission or absorption if $\Delta E=0$,
i.e. the final and the initial states coincide.

It is useful to distinguish the two cases. 
\begin{itemize}
\item In the emission case $E_{\mathrm{fin}}<E_{\mathrm{in}}$, so $E_{\mathrm{ph}}<\Delta E$. 
\item In the absorption case $E_{\mathrm{fin}}>E_{\mathrm{in}}$, so $E_{\mathrm{ph}}>\Delta E$. 
\end{itemize}
We can explain these two results as follows. 
\begin{itemize}
\item In the emission case, part of the energy gap $\Delta E$, which is
a sort of energy ``at disposal'', is taken by the recoiling atom. 
\item In the absorption case, the photon has to supply an extra energy in
order to make the atom recoil after the absorption. 
\end{itemize}
Notice that the photon energy is different between the emission and
the absorption case. In both cases, $E_{\mathrm{ph}}\neq\Delta E$,
as opposed to Bohr model prediction.

Let us evaluate the magnitude of the corrections to Ritz-Rydberg formula,
giving a numerical estimate for transitions between $E_{1}$ and $E_{2}$
for a hydrogen atom at rest. 
\[
\Delta E=E_{2}-E_{1}=\frac{3}{8}\mu c^{2}\alpha^{2}=\unit[10.198\,719\,16]{eV}
\]
\[
E_{\mathrm{em}}=\frac{2Mc^{2}-\frac{5}{8}\mu c^{2}\alpha^{2}}{2\left(Mc^{2}-\frac{1}{8}\mu c^{2}\alpha^{2}\right)}\Delta E=\unit[10.198\,719\,11]{eV}
\]
\[
E_{\mathrm{abs}}=\frac{2Mc^{2}-\frac{5}{8}\mu c^{2}\alpha^{2}}{2\left(Mc^{2}-\frac{1}{2}\mu c^{2}\alpha^{2}\right)}\Delta E=\unit[10.198\,719\,22]{eV}
\]
As one can see, the differences are very small ($\sim\unit[10^{-8}]{eV}$),
so they are extremely difficult to detect with an experimental apparatus.
In addition, probing orders of magnitude of about $\unit[10^{-8}]{eV}$,
we have also to take into account fine and hyperfine corrections,
so this preliminary analysis is largely invalidated at these energies.
Nevertheless we can conclude that, neglecting fine and hyperfine structure,
empirical observations are in agreement with Bohr model up to an excellent
degree of approximation. However, those tiny differences played a
fundamental role in the paradox we have examined.

The formula in eq.~\eqref{eq:summing up} has also an interesting
feature. In classical mechanics, adding a constant term $a$ to the
hydrogen atom Hamiltonian does not change the dynamics of the system.
This is a sort of ``gauge invariance''. This fact prevents us from
knowing what constant $a$ we have chosen to define the Hamiltonian.

In quantum mechanics, adding a constant term $a$ to the Hamiltonian
operator shifts its eigenvalues of a quantity $a$. Since one of the
probes of energy levels of atoms are emitted and absorbed photons,
knowing photon energies allows us to infer atomic energy levels. According
to Bohr model, $E_{\mathrm{ph}}=\Delta E$, so a hypothetical shift
in the energy levels is not detectable. Again, this fact guarantees
a sort of ``gauge invariance'' and prevents us from knowing the
choice of the constant term $a$.

Instead, if in eq.~\eqref{eq:summing up} we shift the energy levels
of a quantity $a$, the photon energy does \emph{not} remain invariant.
Therefore, in principle, from eq.~\eqref{eq:summing up} it would
be possible to infer the constant added to the Hamiltonian operator,
breaking classical ``gauge invariance''.

When the atom is moving, eq.~\eqref{eq:summing up} holds no more.
We have a slightly more complex formula, given by 
\[
E_{\mathrm{ph,moving}}=\frac{1}{\gamma\left(1-\beta\cos\vartheta\right)}E_{\mathrm{ph,rest}},
\]
where $E_{\mathrm{ph,rest}}$ is given by eq.~\eqref{eq:summing up}
and $\vartheta$ is the angle between the atomic motion and the direction
of the emitted or absorbed photon. In general, $E_{\mathrm{ph,moving}}\neq E_{\mathrm{ph,rest}}$,
due to the presence of the kinematic factor $1/\left[\gamma\left(1-\beta\cos\vartheta\right)\right]$.
This factor enables a richer case study of photon energies than the
simple rest situation.

Finally, we have also investigated if photons are really responsible
for atomic transitions between different energy levels, at least from
a special relativistic point of view.

In the emission case, we have proven that a disexcitation cannot take
place without a photon emission: it would imply a 4-momentum conservation
violation. In this case, the 4-momentum conservation method has enabled
us to obtain this fundamental result.

Similarly, one would expect that atomic excitations are impossible
without photon absorption. The first conceptual problem has been how
to characterize the fact that the photon has not been absorbed by
the atom. From an operative point of view, we can see that the photon
has not been absorbed if we find the photon somewhere after the interaction
with the atom by means of an appropriate detector. From a theoretical
point of view, this operative layout is given by the 4-momentum conservation
method, in which we treat only the initial and final states, abandoning
the proposal of a detailed description of the interaction between
the atom and the photon.

Nevertheless, the 4-momentum conservation method is able to exclude
atomic excitation only for photon energies which are lower than the
absorption energy. This method does not rule out atomic excitations
for photon energies which are higher than the absorption energy. Actually,
the matter is that the 4-momentum conservation method does not examine
the actual interaction between the atom and the photon. In this way,
nobody assures us that the photon detected in the final state has
not undertaken an absorption-emission process as described in section~\ref{sec:When-the-photon}.
Therefore, a photon detected in the final state does not grant us
that no photon has ever been absorbed by the atom, whence the 4-momentum
conservation formalism is not well suited to describe the situation
of absorption without photons.
\begin{acknowledgments}
I am particularly grateful to Dr.\,Jean-Pierre Zendri, because this
paradox was inspired by one of his lectures on laser cooling.

Special thanks to Prof.\,Francesco Sorge for his suggestions during
the writing of this article.
\end{acknowledgments}
\appendix

\section{Some calculations\label{sec:Some-calculations}}

In this appendix we present a more mathematically detailed treatment
of the physical situation described in section~\ref{sec:When-the-photon}.

If in the laboratory frame $K$ the atom is at rest in its energy
level $E_{n}$, and a photon with energy $E_{0}$ is moving along
the $x$ axis, their 4-momenta before interaction are 
\[
p_{\mathrm{a}}^{\mu}=\left(\begin{array}{c}
m_{n}c\\
0\\
0\\
0
\end{array}\right)\qquad p_{0}^{\mu}=\left(\begin{array}{c}
E_{0}/c\\
E_{0}/c\\
0\\
0
\end{array}\right),
\]
where $p_{\mathrm{a}}^{\mu}$ is the atom 4-momentum and $p_{0}^{\mu}$
is the photon 4-momentum. After interaction, if we suppose that the
atom jumps to $E_{m}$, we have 
\[
\widetilde{p}_{\mathrm{a}}^{\mu}=\left(\begin{array}{c}
\gamma m_{m}c\\
p_{\mathrm{a}}\vec{n}_{\mathrm{a}}
\end{array}\right)\qquad\widetilde{p}_{0}^{\mu}=\left(\begin{array}{c}
E_{0,\mathrm{fin}}/c\\
\left(E_{0,\mathrm{fin}}/c\right)\vec{n}_{0}
\end{array}\right),
\]
with $\vec{n}_{\mathrm{a}}$ and $\vec{n}_{0}$ two unit vectors.

Let $\varphi$ be the angle between $\vec{n}_{0}$ and the $x$ axis.
Computing Mandelstam $t$ invariant before and after interaction,
and equating the two expressions, one has 
\begin{equation}
-2\frac{E_{0}E_{0,\mathrm{fin}}}{c^{2}}+2\frac{E_{0}E_{0,\mathrm{fin}}}{c^{2}}\cos\varphi=m_{m}^{2}c^{2}+m_{n}^{2}c^{2}-2\gamma m_{n}m_{m}c^{2}.\label{eq:t invariant}
\end{equation}
From energy conservation 
\begin{equation}
\gamma m_{m}c^{2}=m_{n}c^{2}+E_{0}-E_{0,\mathrm{fin}};\label{eq:energy conservation scattering}
\end{equation}
thus, substituting this expression of $\gamma m_{m}c^{2}$ into eq.~\eqref{eq:t invariant},
we have, after some passages, (cf. eq.~\eqref{eq:scattered photon energy})
\begin{equation}
E_{0,\mathrm{fin}}=\frac{\left(m_{n}^{2}-m_{m}^{2}\right)c^{2}+2m_{n}E_{0}}{2\left[E_{0}\left(1-\cos\varphi\right)+m_{n}c^{2}\right]}c^{2}.\label{eq:photon energy scattering}
\end{equation}

Clearly it must be $E_{0,\mathrm{fin}}>0$. The denominator is always
positive, in fact 
\[
m_{n}c^{2}>-E_{0}\left(1-\cos\varphi\right),
\]
and the right-hand side is always non-positive. Hence, $E_{0,\mathrm{fin}}>0$
is equivalent to 
\[
\left(m_{n}^{2}-m_{m}^{2}\right)c^{2}+2m_{n}E_{0}>0,
\]
which yields $E_{0}>E_{\mathrm{abs}}$, where $E_{\mathrm{abs}}$
is given by $E_{\mathrm{abs}}=\left(m_{m}^{2}-m_{n}^{2}\right)c^{2}/\left(2m_{n}\right)$,
which corresponds to eq.~\eqref{eq:absorbed photon energy levels}.

Therefore, the case $E_{0}<E_{\mathrm{abs}}$ is ruled out, because
it would imply $E_{0,\mathrm{fin}}<0$.

In addition we have to check whether $\gamma\geq1$ in eq.~\eqref{eq:energy conservation scattering},
otherwise we have an unphysical situation. Since $\gamma$ is given
by (cf. eq.~\eqref{eq:energy conservation scattering}) 
\[
\gamma=\frac{m_{n}c^{2}+E_{0}-E_{0,\mathrm{fin}}}{m_{m}c^{2}},
\]
we must have 
\begin{equation}
E_{0,\mathrm{fin}}\leq E_{0}-\left(m_{m}-m_{n}\right)c^{2}.\label{eq:second fundamental inequality}
\end{equation}
Let us impose inequality~\eqref{eq:second fundamental inequality}
to the result of eq.~\eqref{eq:photon energy scattering}. We find
an inequality $E_{0}$ has to fulfill. 
\[
2E_{0}^{2}\left(1-\cos\varphi\right)-2E_{0}c^{2}\left(m_{m}-m_{n}\right)\left(1-\cos\varphi\right)+\left(m_{m}-m_{n}\right)^{2}c^{4}\geq0
\]
Its reduced discriminant is 
\[
\frac{\Delta}{4}=-\left(m_{m}-m_{n}\right)^{2}c^{4}\sin^{2}\varphi,
\]
which is always non-positive, so inequality~\eqref{eq:second fundamental inequality}
is always fulfilled.

\end{document}